\documentclass[sigconf]{acmart}

\usepackage{graphicx}

\usepackage{lscape}
\usepackage{xcolor}
\usepackage{lmodern}
\usepackage{eurosym}
\usepackage{url}
\usepackage{tabularx} 
\usepackage[caption = false]{subfig}
\usepackage{tabularx}
\usepackage{multirow}
\usepackage{hhline}

\newcolumntype{L}{@{}l@{}} 

\copyrightyear{2022} 
\acmYear{2022} 
\setcopyright{rightsretained} 
\acmConference[CHASE'22 ]{15th International Conference on Cooperative and Human Aspects of Software Engineering}{May 21--29, 2022}{Pittsburgh, PA, USA}
\acmBooktitle{15th International Conference on Cooperative and Human Aspects of Software Engineering (CHASE'22 ), May 21--29, 2022, Pittsburgh, PA, USA}
\acmDOI{10.1145/3528579.3529173}
\acmISBN{978-1-4503-9342-3/22/05}

\begin{document}

\title{Problem reports and team maturity in agile \\automotive software development}

\author{Lucas Gren}
\affiliation{%
  \institution{Volvo Cars and Chalmers $|$ University of Gothenburg}
    \city{Gothenburg}
  \country{Sweden}}
  \affiliation{%
  \institution{Blekinge Institute of Technology}
  \city{Karlskrona}
  \country{Sweden}}
\email{lucas.gren@lucasgren.com}

\author{Martin Shepperd}
\affiliation{%
  \institution{Chalmers $|$ University of Gothenburg}
    \city{Gothenburg}
  \country{Sweden}}
  \affiliation{%
  \institution{Brunel University}
  \city{London}
  \country{UK}}
\email{martin.shepperd@brunel.ac.uk}

\begin{abstract}
\textbf{Background:} Volvo Cars is pioneering an agile transformation on a large scale in the automotive industry. Social psychological aspects of automotive software development are an under-researched area in general. Few studies on team maturity or group dynamics can be found specifically in the automotive software engineering domain.

\textbf{Objective:} This study is intended as an initial step to fill that gap by investigating the connection between issues and problem reports and team maturity.

\textbf{Method:} We conducted a quantitative study with 84 participants from 14 teams and qualitatively validated the result with the Release Train Engineer having an overview of all the participating teams.

\textbf{Results:} We find that the more mature a team is, the faster they seem to resolve issues as provided through external feedback, at least in the two initial team maturity stages.

\textbf{Conclusion:} This study suggests that working on team dynamics might increase productivity in modern automotive software development departments, but this needs further investigation.
\end{abstract}



\maketitle

\section{Introduction}
An agile approach to projects has become extremely popular both within \citep{dingsoyr20121213}, and outside software development \citep{pajares2017project}. In contrast to traditional plan-driven development (a waterfall process), this paradigm has been given the name \emph{agile}, which literally means to be able to move quickly, easily and be flexible \citep{williams}. 
Feasible solutions to customer needs should be developed in close customer collaboration, rather than being subject to the uncertainties of large up-front design and planning \citep{isleanagile}. With the explosion of software in the car, this complexity has forced automotive companies to, not just focus on customer value, but to also to rely on teams, and teams of teams, to develop modern cars with around 150 Million lines of code. This implies that the social aspect of group dynamics need to be understood in order to scale the development projects. The car projects also differ in that the agile approaches are used across both software, hybrid, and pure hardware teams. This study, though, focuses on the software teams in this quite different context since the $\approx 800$ teams essentially build one single product together.

Research on automotive software engineering has mainly focused on testing and building software models \citep{haghighatkhah2017automotive}. We have not been able to find many studies on organizational psychology in general except studies on ``human factors'' associated with individuals (e.g., \citet{maro2017challenges}). We have not been able to find any studies on social psychology in general, or team maturity in particular, in automotive software development. 



Some studies have been conducted on self-organization in general agile software development (i.e., not automotive) teams in a broader software engineering context, also referred to as team autonomy. \citet{moe2008} concluded that one of the barriers to agile team autonomy was a lack of system support for teams and reduced external autonomy. 
More recently, \citet{hodgson2013controlling} investigated how teams adopt agile practices over time. They conclude that some teams do a ``big bang'' while other implement agile practices one-by-one. They saw a trend that team members without much previous agile knowledge prefer the latter while experienced agile practitioners prefer the former. Volvo Cars has gone through an agile transformation in the last five years and there is a strong push towards providing teams with the mandate to lead themselves to a larger extent, i.e., a ``big bang'' on a very large scale impacting close to 800 teams. 

\citet{hoda2011supporting} showed that senior management have a key role in enabling team agility. That there is a challenge in fitting the agile approach into traditional management in a software engineering context was also shown by \citet{hodgson2013controlling}. In more general leadership research, there are traits that a good leader must have, but on top of that, good leaders in all fields adapt their leadership style dynamically to the situation \citep{2009tei}. Leadership could also be seen as a function of group action \citep{von1986leadership} and not a role, which makes it possible to share or change over time. This means that there is no best leadership style since many different styles are needed depending on the context, the group, and the people involved. In the agile space, e.g., servant leadership is advocated as the foundation for leading, but the definition of what that is remains vague \citep{parris2013systematic}. Leadership is a difficult construct to research since it happens in complex systems and is influenced by a vast number of factors.

\citet{grenjss2} found initial indications confirming that the definition of agile teams overlaps with what is meant by a mature group in social psychology \citep{wheelan}. Agile teams are dependent on also being mature in terms of collaborative behaviur plus having navigated through the initial stages of group development. Mature teams are also known to collaborate better with other teams \citep{wheelandev}, which is of interest when scaling the agile ways of working at Volvo Cars. We define team maturity in this study as the degree to which a team has navigated through the group development stages according to \citet{wheelandev}. In Wheelan's Integrated Model of Group Development (or IMGD), small groups (or teams) start in the Dependency and Inclusion stage where team members are tentative and polite in their communication and need to build an initial level of psychological safety. Team members do not have a common mental model of the group goal nor do they know the real competencies of the other team members. When higher levels of psychological safety start to emerge, the team transitions into the second phase, Counter-Dependency and Fight. This is a conflict stage but is absolutely necessary to determine who can contribute to what and create a shared mental model of the team's purpose. After the second stage, the team can organize work better and set productive group norms. With higher levels of trust and a continuous improvements of work organization, the team can reach stage four, which is the Work and Productivity stage \citep{wheelandev}.

Leadership, self-organization, and agility can thus be mapped onto the team development stages \citep{grenjss2}. As the development cycle moves along from a psychological perspective, all members of the team (including the managers or leaders) must change their behavior \citep{wheelandev}. Team maturity, as measured by the Group Development Questionnaire (GDQ), has been shown to be correlated to measurements of productivity also in software engineering \citep{al2018connections}. In this study, we used a recently developed short version of the GDQ comprising 13 items instead of 60 \citep{gren2020gdqs}, which makes it practically more useful considering the time availability of teams.

If the results on the connection between team maturity and team productivity from social psychology \citep{wheelandev} apply to automotive software development, aspects of productivity should also be connected to measurements of team maturity in that context. One quite objective proxy for productivity for these Volvo Cars teams was their issue management, i.e., how long the teams take to fix issues reported back from their deliveries. This metric is external to the team, which makes it in many ways more appropriate than e.g., flow of backlog items or self-assessed productivity.

\section{Method}
We distributed the short version of the GDQ, (the GDQS \citep{gren2020gdqs}) to 14 teams engaged in software-only development at Volvo Cars. The teams also logged the number of issues they have with the delivered software in the same month. Issues could be for example bugs or other types of change requests. The number of issues can accumulate since many of them remain unaddressed so it may not a good measurement of the current team performance. We, therefore, instead looked at the average age of issues in the same month as the teams filled out the GDQS. The average is sensitive to outliers but Volvo Cars chose that metric as a KPI since all issues need to be resolved. 82 team members answered the GDQS from all the 14 teams. Since this is an average of responses per team and the teams are usually from 3 to 12 individuals, we estimate this response rate to be around 80\%.

We describe the data and its patterns or lack thereof and avoid using $p$ values for null hypothesis significance testing in accordance with \citet{mcshane2019abandon}. We instead use an approach based on Bayesian Data Analysis (BDA), i.e., we will plot the re-sampled (using Markov-Chain Monte Carlo) distribution of all our parameters and be explicit about our assumptions (see supplementary material). For a description and tutorial applied to software engineering see \citet{furia2019bayesian}.

Note all the shared data in this paper, plots and supplementary material were transformed using arbitrary linear transformations. This was done so as to not reveal the absolute values of issues and team maturity, since that was deemed as sensitive information by the company. The average ages of issues where downloaded from the Volvo Cars instance of Jira and the team maturity data was collected through the company's own survey tool. 

In the second step, we interviewed the Release Train Engineer (RTE), which is a role that has an overview of this whole set of teams. The interviewee was asked to describe the situation around the data collection and also critically assess the connection between team maturity and the teams' capacity to resolve reported issues. 

Based on previous studies of team maturity and aspects of productivity (see Section~1), we hypothesize that the first two stages (Stage 1 and 2) should be positively connected to the average age of issues, i.e., the more collaborative aspects to solve in these stages, the longer it would take to solve reported issues. We would also hypothesize that the latter stages (Stage 3 and 4) would have the opposite relationship to reported issues, since higher values on these two scales imply more maturity (e.g., higher levels of trust etc.).


\section{Results}
\subsection{Quantitative results}
\begin{figure}
\subfloat{\includegraphics[width = 1.75in]{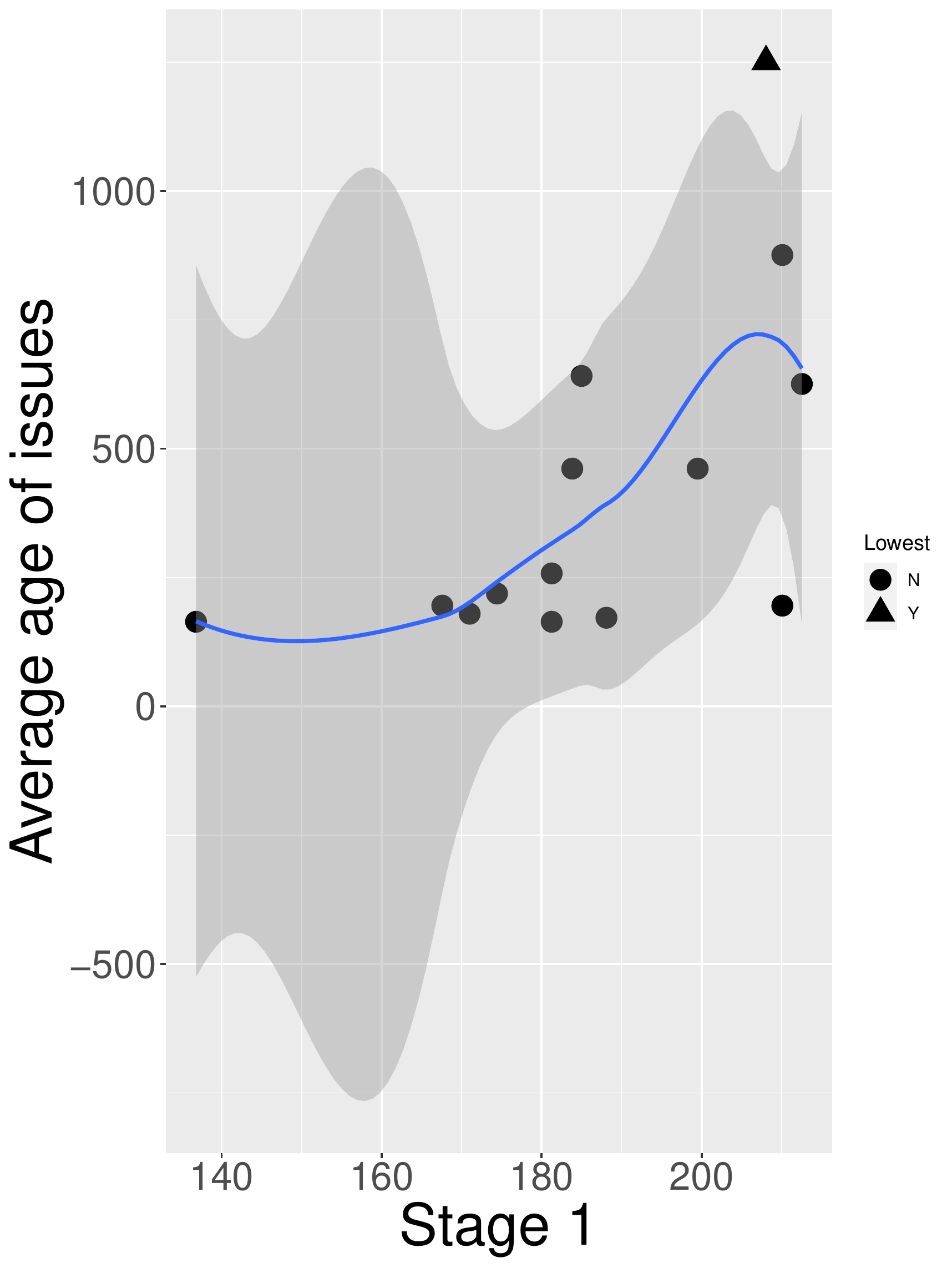}} 
\subfloat{\includegraphics[width = 1.75in]{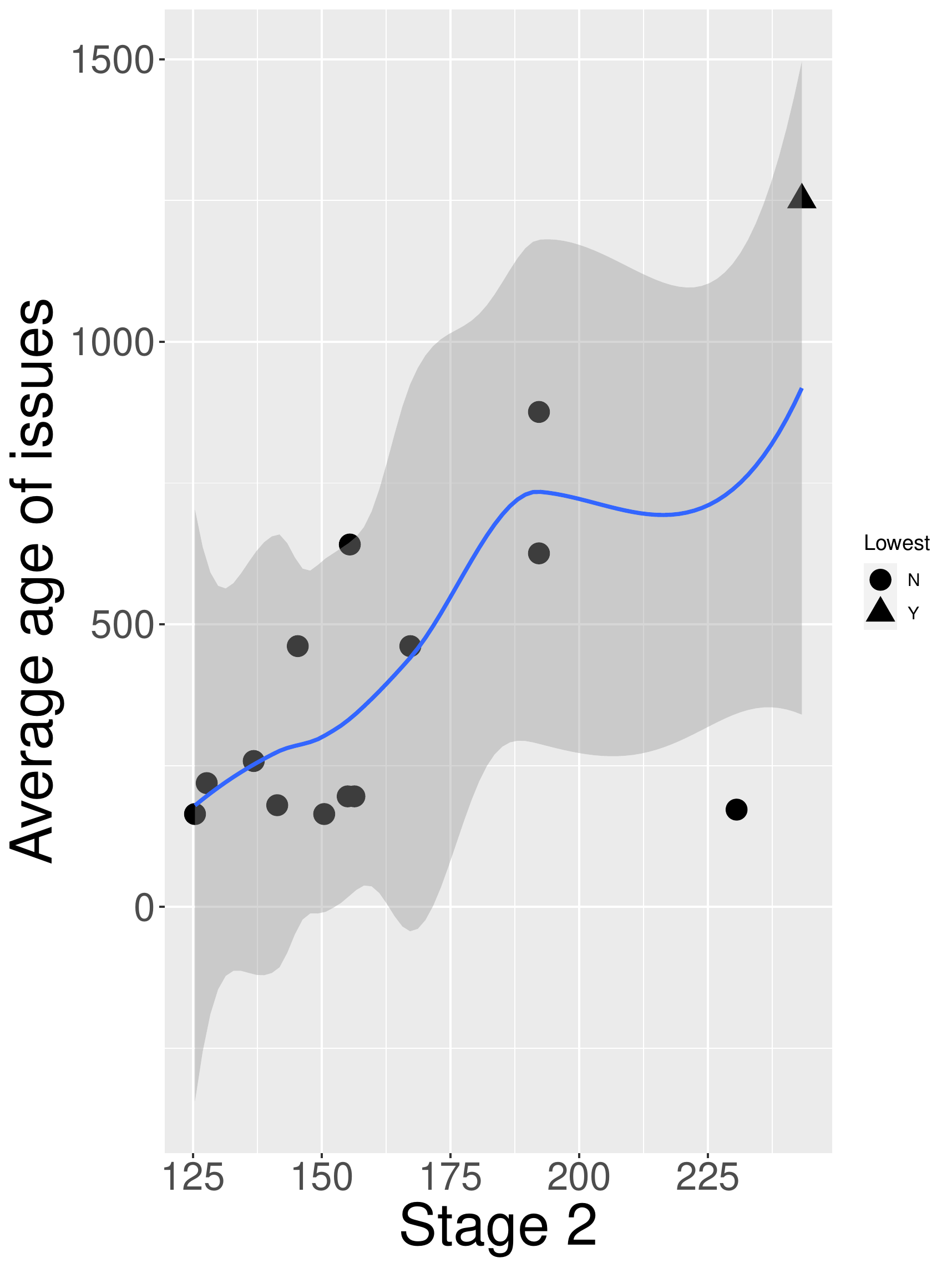}}\\
\subfloat{\includegraphics[width = 1.75in]{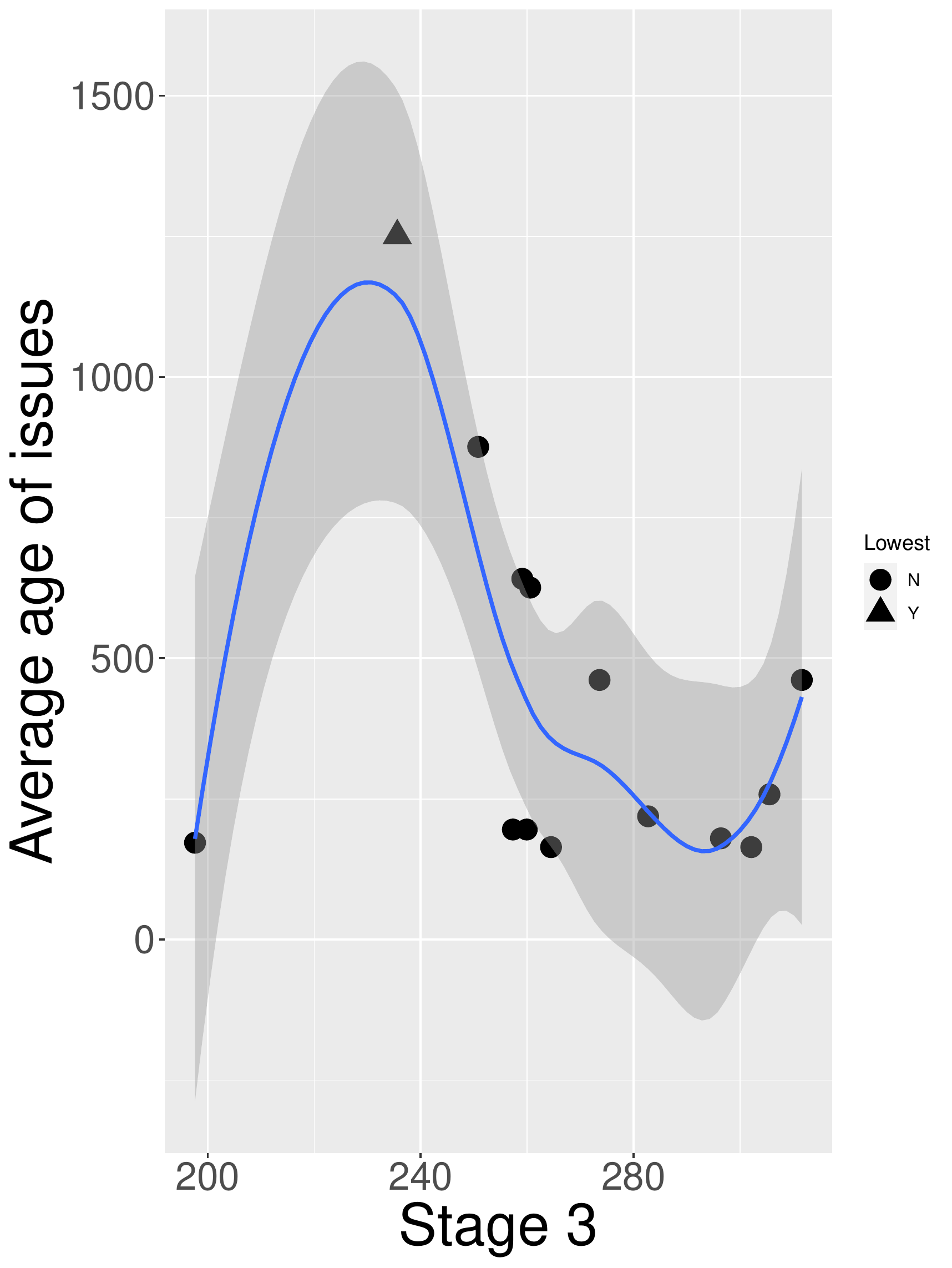}}
\subfloat{\includegraphics[width = 1.75in]{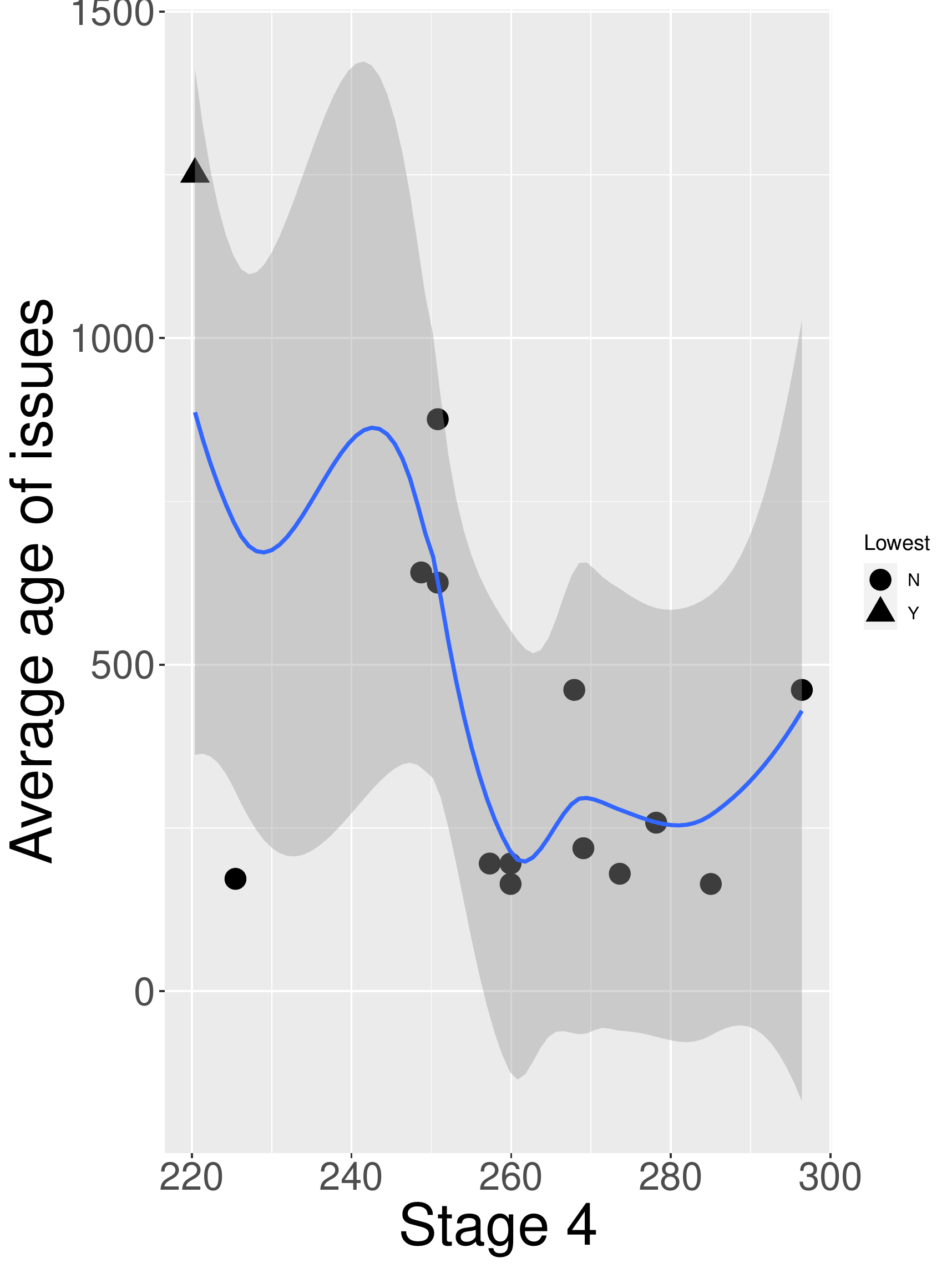}}
\caption{Scatter plots with each with one of the developmental stages of team maturity as independent variable (the x axis) and the average age of issues as dependent variable (the y axis).}
\label{corrplots}
\end{figure}

\begin{figure}
\subfloat[Stage 1]{\includegraphics[width = 1.75in]{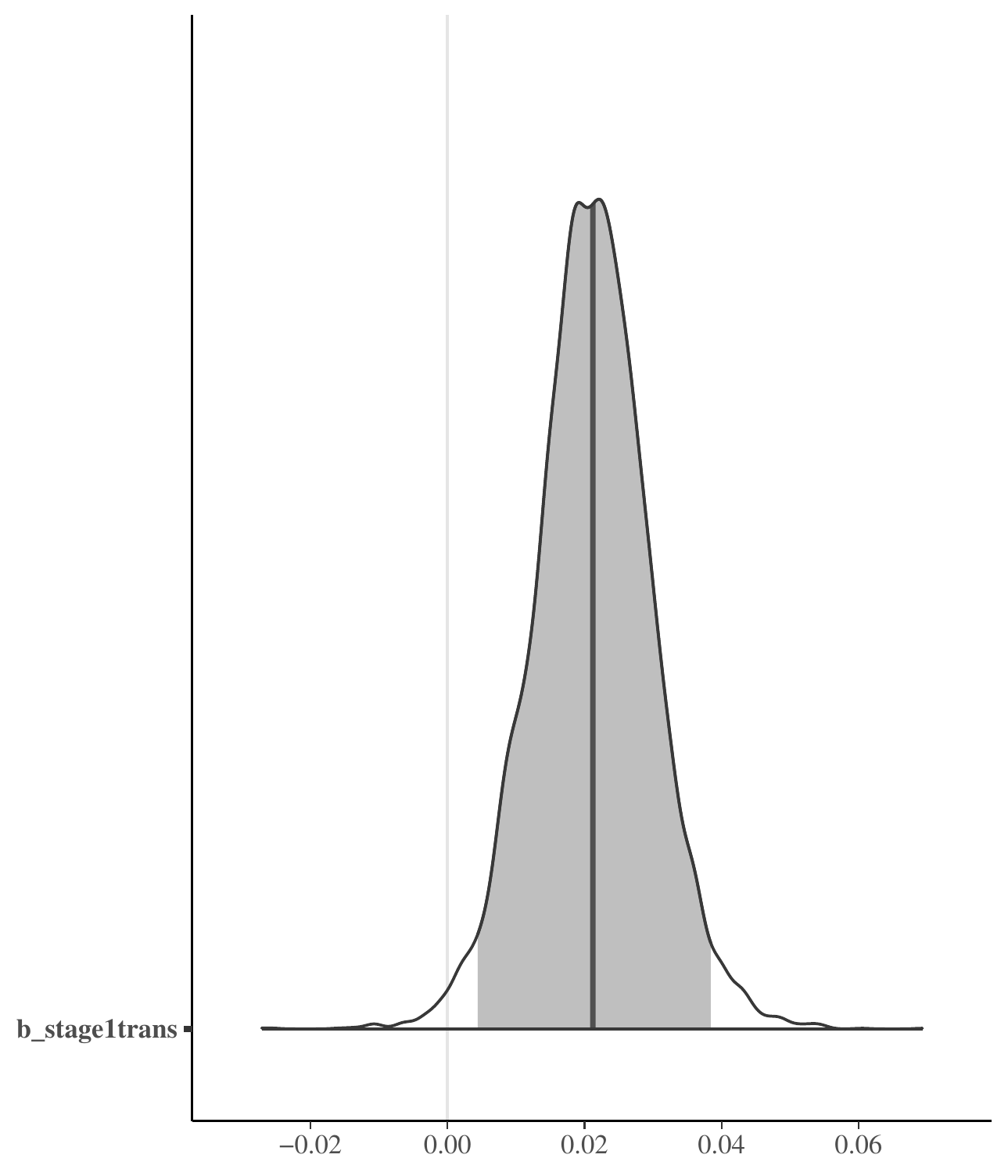}} 
\subfloat[Stage 2]{\includegraphics[width = 1.75in]{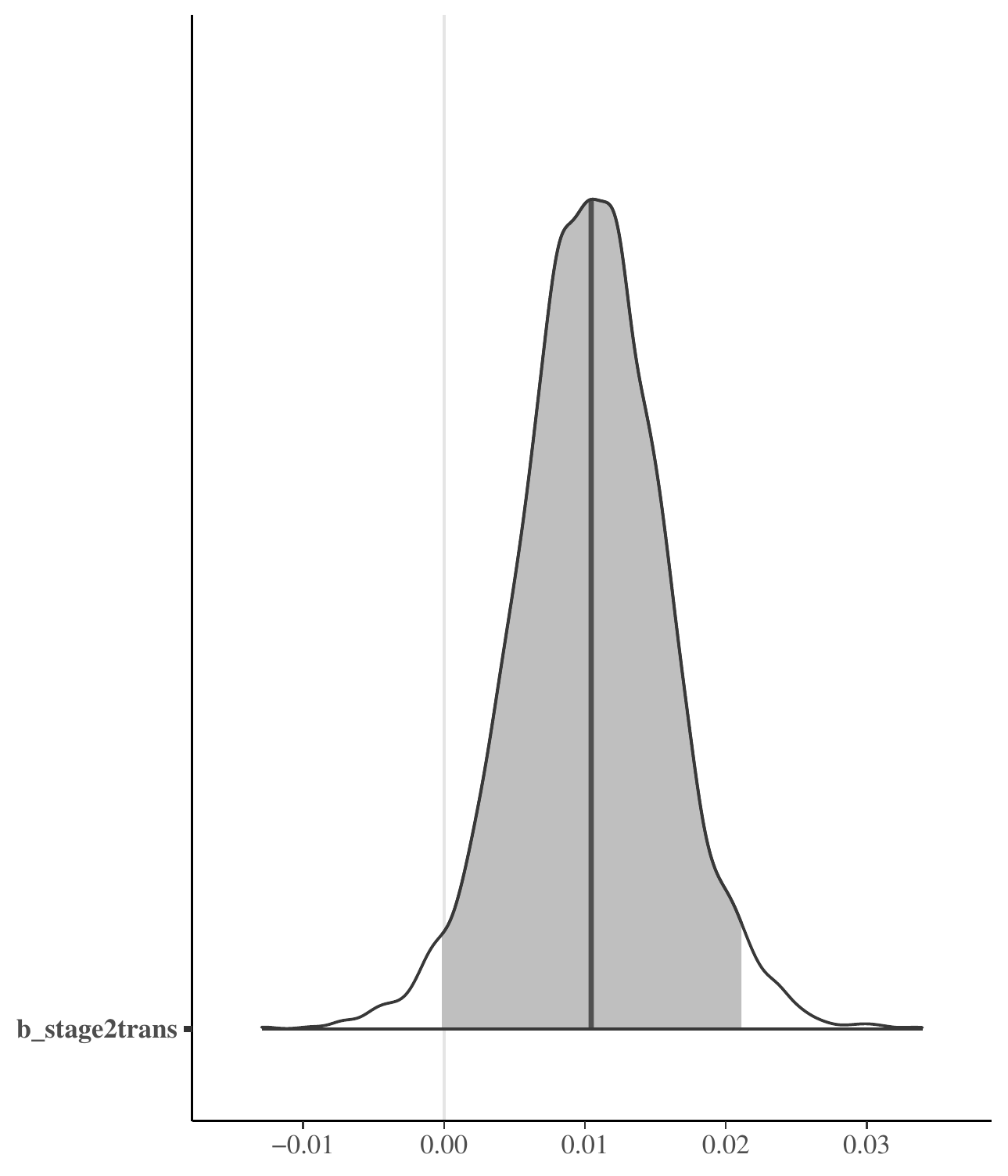}}\\
\subfloat[Stage 3]{\includegraphics[width = 1.75in]{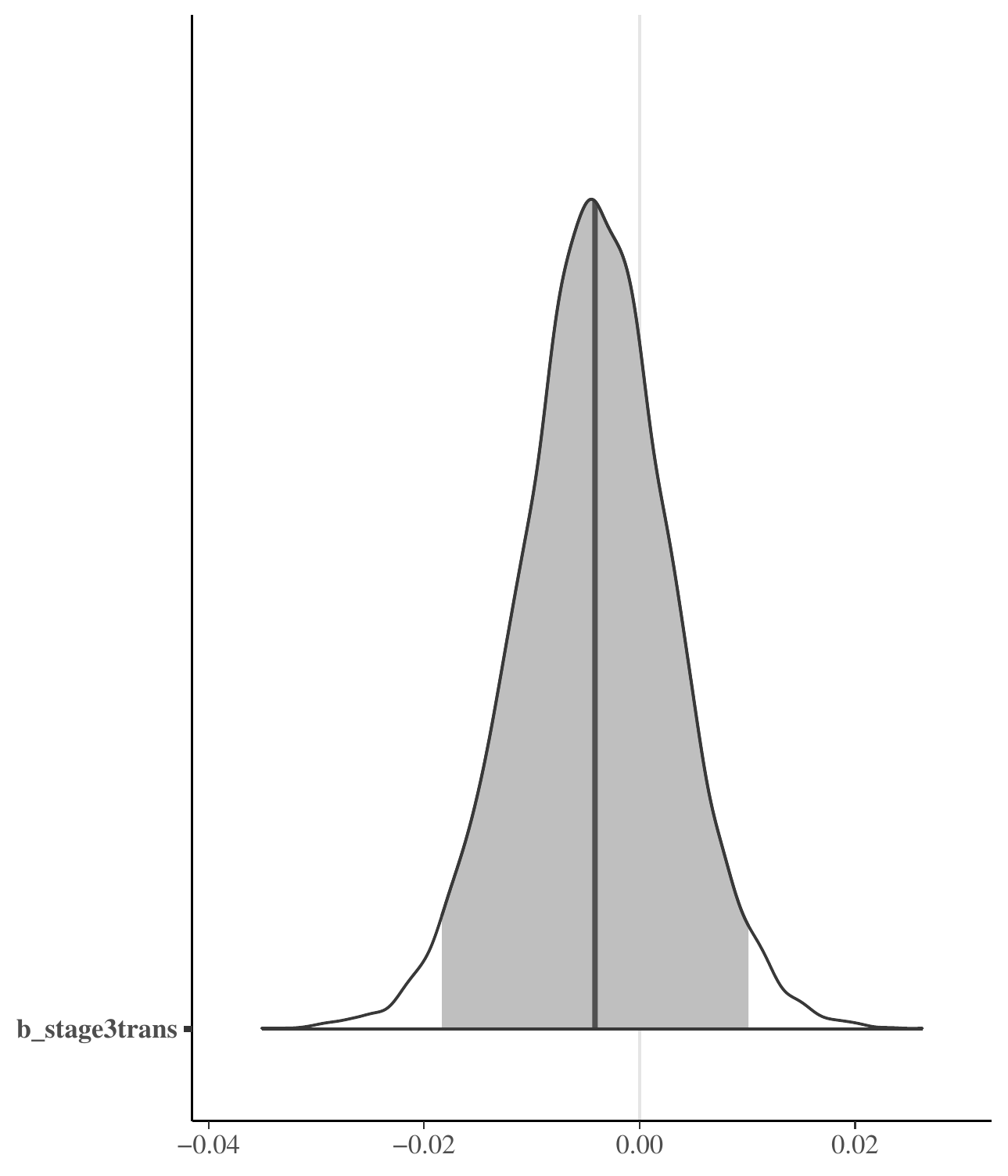}}
\subfloat[Stage 4]{\includegraphics[width = 1.75in]{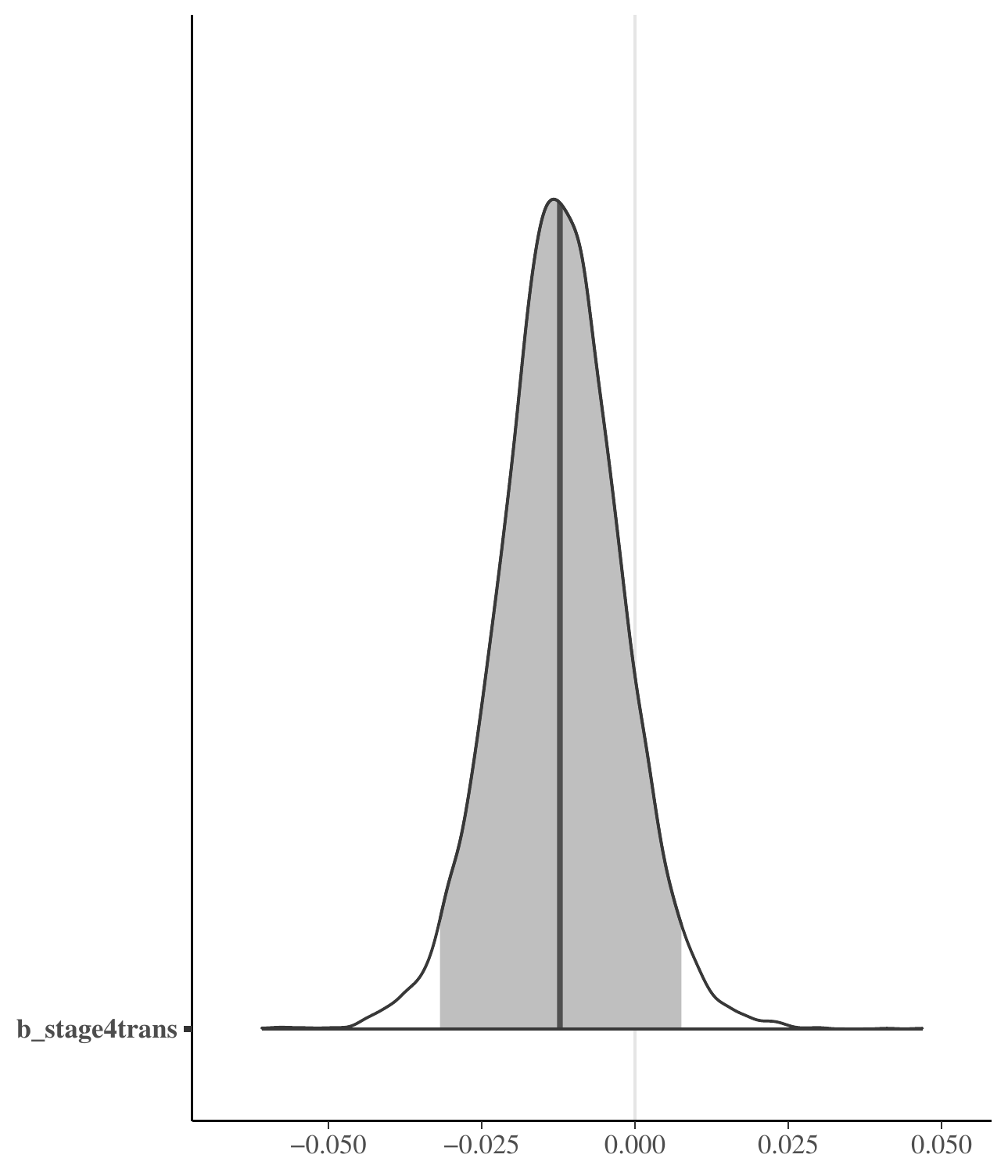}}
\caption{Re-sample distributions of four separate models, each with one of the developmental stages of team maturity as independent variable and the average age of issues as dependent variable.}
\label{mcmc}
\end{figure}

The result shown in Fig.~\ref{corrplots} indicates that the direction of the lines are in the hypothesized direction. However, Fig.~\ref{mcmc} shows that only the distributions of stages 1 and 2 are distinct from zero with close to a 95\% credible interval. 

As mentioned, we do not want to use classical NHST inference, which means that we see this quantitative result as indicative, but further studies are needed. We would like to, instead, highlight the importance of the qualitative data below.

In summary, when the teams are newly formed (or less mature) or in the conflict stage of creating shared mental models of the team goals and figuring out who can really contribute with what in order to fulfill these goals, the average age of issues is higher for those teams.

\subsection{Qualitative validation}
Below is a summary of what the RTE commented upon in the interview. 

During the time of the data collection, the teams were focusing on removing issues that were never to be dealt with. The team developed their software in-house and wanted to remove issues they had inherited from suppliers. The teams were working on two different platforms, one old and one new. Different teams were working on the different platforms. The teams working on the old platforms were described as having lower motivation for their work compared to the teams working on the new platform. The teams that worked on the old platform would have preferred to work on the new platform instead. The teams working on the new platform used to even be located at a different municipality. 

The performance and collaboration of the teams were described as very different. Regarding their reported issues, some teams could solve those internally while other teams were highly dependent on suppliers. For some teams, solving issues was a complex task with many steps. 

The teams working on the old platform had closed a high number of issues, other teams received reports of defects that were incorrectly perceived as being errors. 

The teams working on the old platform were assessed as having lower team maturity compared to the ones working on the new platform. They closed a lot of issues that they realized would never be dealt with. Fixing issues from problem reports was a part of all the teams' daily work. 

As a form of validation, the team with the highest values on the y axis in Fig.~\ref{corrplots}, was the team the interviewee identified as the least mature team (denoted as a triangle). 

\section{Discussion}
The results of this study suggest that the automotive software development having transformed into agile ways of working, might be dependent on team maturity in order to deliver value in the shape of fixing issues in their delivery. Since the team as a unit of analysis lies at the heart of an agile organization, research on automotive agile software development should also use theories and models from social psychology in order to understand the mechanisms at play.

On the necessary journey towards being more responsive to change, Volvo Cars has been on an agile transformation for the last five years. Some parts of the company has gotten farther than others at the time of this research, but the vehicle software and electronics department were the forerunner in the transformation. 

Now that the automotive industry has software as a key to gaining a competitive advantage, research on software engineering adapted to the automotive industry is needed and conducted to understand this somewhat specific context. The psychological aspects of automotive software engineering has, though, not been much in focus and this study is a tiny step in beginning to fill that gap. Research so far has focused on testing and building software models \citep{haghighatkhah2017automotive} and the human factors investigated have been associated with individual psychology (e.g., \citet{maro2017challenges}). Since the aspects of agile team autonomy now also surfaces in the automotive software development, we need to investigate this context further also with regards to self-organization. \citet{moe2008} concluded, among other things, that one of the barriers of agile team autonomy was a lack of system for team support. This is an aspect identified in the agile transformation at Volvo Cars as well and the participating teams (and many others) were trained for at least three hours in Wheelan's Integrated Model of Group Development. They were also given access to a toolbox connected to the different stages after having filled out the GDQS, which hopefully helped them maturing further as a team, but this remains to be investigated.

\subsection{Practical implications}
We have assumed that the team maturity would be connected to aspects of productivity in the automotive software development at Volvo Cars, but this current study is the first one showing data on this actually being the case. If the results of this study would hold for automotive software engineering in general, it has a multitude of implications for developing such software. Many of the challenges go hand-in-hand with the agile transformation, but also provides way of getting teams to mature faster. One of these implications is to focus even more on training leaders and managers (in addition to training teams, of course) in adapting their leadership style dynamically to the team maturity level \citep{2009tei}. A team will not self-organize on day one because the team members are unlikely to have met before.  Also there are the leadership challenges of designing teams well and assuring the emergence of good work principles and norms \citep{magdalucasxp}, instead of micromanaging technical details. Agile leadership has been described as ``servant leadership,'' which remains vague \citep{parris2013systematic}. But a key aspect of thinking about servant leadership is that it challenges many managers' mental model of leadership being based on hierarchical power, however, agile leadership need to be a servant in a way that serves the teams with what they actually need right now. That could also be to confront behavior that does not build psychological safety, as an example \citep{agileleadicse}.

This current study is in line with the result by \citet{al2018connections} that team maturity as measured by the GDQ is connected to productivity aspect of software development, but extends it into the automotive software development domain. A very practical implication for the automotive industry is to measure team maturity in a short validated survey, such as the GDQS \citep{gren2020gdqs}, train both managers, agile leaders, and team members in team development, by for example using the Integrated Model of Group Development \citep{wheelan} in order to increase the speed of fixing issues in the teams' deliveries. We infer this causality by logic and the qualitative data, since it does not make sense to hypothesize that fixing issues fast would lead to the team maturing, however, the mediating role of team maturity between team building interventions and the speed of fixing issues, of course needs much more research. 

\subsection{Threats to validity}
We use relevant parts of the checklist created by \citet{gren2018standards} for behavioral software engineering research.

\paragraph{Reliability}
Both the internal consistency and stability have been checked multiple times in the development of the GDQ \citep{wheelan} and GDQS \citep{gren2020gdqs}. The lack of a formal definition of what becomes an issue in the problem reports at Volvo Cars is a threat to the reliability of us actually measuring the average age of issues the teams have. Further studies should explicitly ask for such a definition, or help define it, if it does not exist. We do not know how extreme values might have affected this average that Volvo Cars chose to use as a KPI. Perhaps the median age of issues might have been more reliable. 

\paragraph{Construct validity}
The constructs under investigation were the four team development scales and the average age of issues. The GDQS is a well-validated scale for measuring team maturity and we only selected teams in a part of Volvo Cars that we know reported and maintained their problem reports continuously. The validity of the used constructs are, therefore, assessed as high.

\paragraph{Internal validity}
We tried to argue for a theoretical reason for causality between team maturity and average age of issues, but we did not measure any other confounding factors. Potential confounds could be type of work, team composition, conformance to processes, individual competencies, etc., and in a social complex adaptive system there could be many more.

\paragraph{External validity}
The small sample of 14 teams and a collection of data from only one part of Volvo Cars are other large threats to this study. We do not know if these teams and the way they work is representative to any teams outside this context. A random sample was not possible in this case since it was hard to find a set of teams that were in a position to continuously report on, and receive feedback from their deliveries. This is an ongoing transformation at Volvo Cars, i.e., to become more data-driven. However, it does seem plausible that they could be applicable to other similar safety-critical software development environments.

\section{Conclusion and Future Work}
We conclude that the average ages of issues in automotive software development teams were associated with teams having higher values on the first two stages of team development. This could mean that mature teams fix issues faster. However, this is just an initial study with a small sample and future studied should sample randomly and also use other productivity measurements. 
Another part of the automotive software development that could be studied in the future is extending the work of \citet{safetylenberg} and further look at how teams deal with safety-critical systems.  

\section*{Supplementary Material}
The Supplementary Material is available on Zenodo\footnote{\url{https://www.doi.org/10.5281/zenodo.6400809}}. It includes an anonymized raw data CSV file and all the R scripts used. 

\section*{Acknowledgments}

Both authors would like to thank the Swedish Research Council funding this work under the auspices of the 2022 Tage Erlander research professorship. Both authors thank Volvo for the provision of the data that made this work possible. This project is an associated project under the Software Center, Gothenburg, Sweden.

\bibliographystyle{ACM-Reference-Format}
\bibliography{ref}


\begin{thebibliography}{00}


\ifx \showCODEN    \undefined \def \showCODEN     #1{\unskip}     \fi
\ifx \showDOI      \undefined \def \showDOI       #1{{\tt DOI:}\penalty0{#1}\ }
  \fi
\ifx \showISBNx    \undefined \def \showISBNx     #1{\unskip}     \fi
\ifx \showISBNxiii \undefined \def \showISBNxiii  #1{\unskip}     \fi
\ifx \showISSN     \undefined \def \showISSN      #1{\unskip}     \fi
\ifx \showLCCN     \undefined \def \showLCCN      #1{\unskip}     \fi
\ifx \shownote     \undefined \def \shownote      #1{#1}          \fi
\ifx \showarticletitle \undefined \def \showarticletitle #1{#1}   \fi
\ifx \showURL      \undefined \def \showURL       #1{#1}          \fi
\providecommand\bibfield[2]{#2}
\providecommand\bibinfo[2]{#2}
\providecommand\natexlab[1]{#1}
\providecommand\showeprint[2][]{arXiv:#2}

\bibitem[\protect\citeauthoryear{Al-Sabbagh and Gren}{Al-Sabbagh and
  Gren}{2018}]%
        {al2018connections}
\bibfield{author}{\bibinfo{person}{K. Al-Sabbagh} {and} \bibinfo{person}{L.
  Gren}.} \bibinfo{year}{2018}\natexlab{}.
\newblock \showarticletitle{The connections between group maturity, software
  development velocity, and planning effectiveness}.
\newblock \bibinfo{journal}{{\em Journal of Software: Evolution and Process\/}}
  \bibinfo{volume}{30}, \bibinfo{number}{1} (\bibinfo{year}{2018}),
  \bibinfo{pages}{e1896}.
\newblock


\bibitem[\protect\citeauthoryear{Dings{\o}yr, Nerur, Balijepally, and
  Moe}{Dings{\o}yr et~al\mbox{.}}{2012}]%
        {dingsoyr20121213}
\bibfield{author}{\bibinfo{person}{T. Dings{\o}yr}, \bibinfo{person}{S. Nerur},
  \bibinfo{person}{V. Balijepally}, {and} \bibinfo{person}{N. Moe}.}
  \bibinfo{year}{2012}\natexlab{}.
\newblock \showarticletitle{A decade of agile methodologies: {T}owards
  explaining agile software development}.
\newblock \bibinfo{journal}{{\em The Journal of Systems and Software\/}}
  \bibinfo{volume}{85} (\bibinfo{year}{2012}), \bibinfo{pages}{1213—--1221}.
\newblock


\bibitem[\protect\citeauthoryear{Furia, Feldt, and Torkar}{Furia
  et~al\mbox{.}}{2019}]%
        {furia2019bayesian}
\bibfield{author}{\bibinfo{person}{C. Furia}, \bibinfo{person}{R. Feldt}, {and}
  \bibinfo{person}{R. Torkar}.} \bibinfo{year}{2019}\natexlab{}.
\newblock \showarticletitle{Bayesian data analysis in empirical software
  engineering research}.
\newblock \bibinfo{journal}{{\em IEEE Transactions on Software Engineering\/}}
  \bibinfo{volume}{47}, \bibinfo{number}{9} (\bibinfo{year}{2019}),
  \bibinfo{pages}{1786--1810}.
\newblock


\bibitem[\protect\citeauthoryear{Gren}{Gren}{2018}]%
        {gren2018standards}
\bibfield{author}{\bibinfo{person}{L. Gren}.} \bibinfo{year}{2018}\natexlab{}.
\newblock \showarticletitle{Standards of validity and the validity of standards
  in behavioral software engineering research: {T}he perspective of
  psychological test theory}. In \bibinfo{booktitle}{{\em International
  Symposium on Empirical Software Engineering and Measurement (ESEM)}}.
  \bibinfo{pages}{1--4}.
\newblock


\bibitem[\protect\citeauthoryear{Gren, Jacobsson, Rydbo, and Lenberg}{Gren
  et~al\mbox{.}}{2020}]%
        {gren2020gdqs}
\bibfield{author}{\bibinfo{person}{L. Gren}, \bibinfo{person}{C. Jacobsson},
  \bibinfo{person}{N. Rydbo}, {and} \bibinfo{person}{P. Lenberg}.}
  \bibinfo{year}{2020}\natexlab{}.
\newblock \bibinfo{title}{The Group Development Questionnaire Short (GDQS)
  Scales: Tiny-Yet-Effective Measures of Team/Small Group Development}.
\newblock   (\bibinfo{date}{Feb} \bibinfo{year}{2020}).
\newblock
\showDOI{%
\url{http://dx.doi.org/10.31234/osf.io/u3p8c}}


\bibitem[\protect\citeauthoryear{Gren and Lindman}{Gren and Lindman}{2020}]%
        {magdalucasxp}
\bibfield{author}{\bibinfo{person}{L. Gren} {and} \bibinfo{person}{M.
  Lindman}.} \bibinfo{year}{2020}\natexlab{}.
\newblock \showarticletitle{What an Agile Leader Does: {T}he Group Dynamics
  Perspective}. In \bibinfo{booktitle}{{\em International Conference on Agile
  Software Development (XP)}}. Springer, \bibinfo{pages}{178--194}.
\newblock


\bibitem[\protect\citeauthoryear{Gren and Ralph}{Gren and Ralph}{2022}]%
        {agileleadicse}
\bibfield{author}{\bibinfo{person}{L. Gren} {and} \bibinfo{person}{P. Ralph}.}
  \bibinfo{year}{2022}\natexlab{}.
\newblock \showarticletitle{What Makes Effective Leadership in Agile Software
  Development Teams?}. In \bibinfo{booktitle}{{\em The 44th International
  Conference on Software Engineering}}. \bibinfo{publisher}{ACM},
  \bibinfo{address}{New York}.
\newblock


\bibitem[\protect\citeauthoryear{Gren, Torkar, and Feldt}{Gren
  et~al\mbox{.}}{2017}]%
        {grenjss2}
\bibfield{author}{\bibinfo{person}{L Gren}, \bibinfo{person}{R Torkar}, {and}
  \bibinfo{person}{R Feldt}.} \bibinfo{year}{2017}\natexlab{}.
\newblock \showarticletitle{Group development and group maturity when building
  agile teams: {A} qualitative and quantitative investigation at eight large
  companies}.
\newblock \bibinfo{journal}{{\em The Journal of Systems and Software\/}}
  \bibinfo{volume}{124} (\bibinfo{year}{2017}), \bibinfo{pages}{104—--119}.
\newblock


\bibitem[\protect\citeauthoryear{Haghighatkhah, Banijamali, Pakanen, Oivo, and
  Kuvaja}{Haghighatkhah et~al\mbox{.}}{2017}]%
        {haghighatkhah2017automotive}
\bibfield{author}{\bibinfo{person}{A. Haghighatkhah}, \bibinfo{person}{A.
  Banijamali}, \bibinfo{person}{O-P. Pakanen}, \bibinfo{person}{M. Oivo}, {and}
  \bibinfo{person}{P. Kuvaja}.} \bibinfo{year}{2017}\natexlab{}.
\newblock \showarticletitle{Automotive software engineering: A systematic
  mapping study}.
\newblock \bibinfo{journal}{{\em Journal of Systems \& Software\/}}
  \bibinfo{volume}{128} (\bibinfo{year}{2017}), \bibinfo{pages}{25--55}.
\newblock


\bibitem[\protect\citeauthoryear{Hoda, Noble, and Marshall}{Hoda
  et~al\mbox{.}}{2011}]%
        {hoda2011supporting}
\bibfield{author}{\bibinfo{person}{R. Hoda}, \bibinfo{person}{J. Noble}, {and}
  \bibinfo{person}{S. Marshall}.} \bibinfo{year}{2011}\natexlab{}.
\newblock \showarticletitle{Supporting self-organizing agile teams}. In
  \bibinfo{booktitle}{{\em International Conference on Agile Software
  Development}}. Springer, \bibinfo{pages}{73--87}.
\newblock


\bibitem[\protect\citeauthoryear{Hodgson and Briand}{Hodgson and
  Briand}{2013}]%
        {hodgson2013controlling}
\bibfield{author}{\bibinfo{person}{D. Hodgson} {and} \bibinfo{person}{L.
  Briand}.} \bibinfo{year}{2013}\natexlab{}.
\newblock \showarticletitle{Controlling the uncontrollable:‘Agile’teams and
  illusions of autonomy in creative work}.
\newblock \bibinfo{journal}{{\em Work, employment and society\/}}
  \bibinfo{volume}{27}, \bibinfo{number}{2} (\bibinfo{year}{2013}),
  \bibinfo{pages}{308--325}.
\newblock


\bibitem[\protect\citeauthoryear{Kozlowski, Watola, Jensen, Kim, and
  Botero}{Kozlowski et~al\mbox{.}}{2009}]%
        {2009tei}
\bibfield{author}{\bibinfo{person}{S. Kozlowski}, \bibinfo{person}{D. Watola},
  \bibinfo{person}{J. Jensen}, \bibinfo{person}{B. Kim}, {and}
  \bibinfo{person}{I. Botero}.} \bibinfo{year}{2009}\natexlab{}.
\newblock \showarticletitle{Developing adaptive teams: {A} theory of dynamic
  team leadership}.
\newblock In \bibinfo{booktitle}{{\em Team effectiveness in complex
  organizations: {C}ross-disciplinary perspectives and approaches}},
  \bibfield{editor}{\bibinfo{person}{E.~Salas}, \bibinfo{person}{G.~Goodwin},
  {and} \bibinfo{person}{S.~Burke}} (Eds.). \bibinfo{publisher}{Routledge},
  \bibinfo{address}{New York}, \bibinfo{pages}{113--155}.
\newblock


\bibitem[\protect\citeauthoryear{Lenberg, Feldt, Wallgren~Tengberg, and
  Gren}{Lenberg et~al\mbox{.}}{2020}]%
        {safetylenberg}
\bibfield{author}{\bibinfo{person}{P. Lenberg}, \bibinfo{person}{R. Feldt},
  \bibinfo{person}{L. Wallgren~Tengberg}, {and} \bibinfo{person}{L. Gren}.}
  \bibinfo{year}{2020}\natexlab{}.
\newblock \showarticletitle{Behavioral Aspects of Safety-Critical Software
  Development}. In \bibinfo{booktitle}{{\em Cooperative and Human Aspects of
  Software Engineering (CHASE)}}. IEEE.
\newblock


\bibitem[\protect\citeauthoryear{Maro, Staron, and Stegh{\"o}fer}{Maro
  et~al\mbox{.}}{2017}]%
        {maro2017challenges}
\bibfield{author}{\bibinfo{person}{S. Maro}, \bibinfo{person}{M. Staron}, {and}
  \bibinfo{person}{J. Stegh{\"o}fer}.} \bibinfo{year}{2017}\natexlab{}.
\newblock \showarticletitle{Challenges of establishing traceability in the
  automotive domain}. In \bibinfo{booktitle}{{\em International Conference on
  Software Quality}}. Springer, \bibinfo{pages}{153--172}.
\newblock


\bibitem[\protect\citeauthoryear{McShane, Gal, Gelman, Robert, and
  Tackett}{McShane et~al\mbox{.}}{2019}]%
        {mcshane2019abandon}
\bibfield{author}{\bibinfo{person}{B. McShane}, \bibinfo{person}{D. Gal},
  \bibinfo{person}{A. Gelman}, \bibinfo{person}{C. Robert}, {and}
  \bibinfo{person}{J. Tackett}.} \bibinfo{year}{2019}\natexlab{}.
\newblock \showarticletitle{Abandon statistical significance}.
\newblock \bibinfo{journal}{{\em The American Statistician\/}}
  \bibinfo{volume}{73}, \bibinfo{number}{sup1} (\bibinfo{year}{2019}),
  \bibinfo{pages}{235--245}.
\newblock


\bibitem[\protect\citeauthoryear{Moe, Dings{\o}yr, and Dyb{\aa}}{Moe
  et~al\mbox{.}}{2008}]%
        {moe2008}
\bibfield{author}{\bibinfo{person}{N. Moe}, \bibinfo{person}{T. Dings{\o}yr},
  {and} \bibinfo{person}{T. Dyb{\aa}}.} \bibinfo{year}{2008}\natexlab{}.
\newblock \showarticletitle{Understanding self-organizing teams in agile
  software development}. In \bibinfo{booktitle}{{\em 19th Australian Conference
  on Software Engineering (ASWEC'08)}}. \bibinfo{pages}{76--85}.
\newblock


\bibitem[\protect\citeauthoryear{Pajares, Poza, Villafa{\~n}ez, and
  L{\'o}pez-Paredes}{Pajares et~al\mbox{.}}{2017}]%
        {pajares2017project}
\bibfield{author}{\bibinfo{person}{J. Pajares}, \bibinfo{person}{D. Poza},
  \bibinfo{person}{F. Villafa{\~n}ez}, {and} \bibinfo{person}{A.
  L{\'o}pez-Paredes}.} \bibinfo{year}{2017}\natexlab{}.
\newblock \showarticletitle{Project Management Methodologies in the Fourth
  Technological Revolution}.
\newblock In \bibinfo{booktitle}{{\em Advances in Management Engineering}},
  \bibfield{editor}{\bibinfo{person}{H.~Cesáreo}} (Ed.).
  \bibinfo{publisher}{Springer}, \bibinfo{pages}{121--144}.
\newblock


\bibitem[\protect\citeauthoryear{Parris and Peachey}{Parris and
  Peachey}{2013}]%
        {parris2013systematic}
\bibfield{author}{\bibinfo{person}{D. Parris} {and} \bibinfo{person}{J.
  Peachey}.} \bibinfo{year}{2013}\natexlab{}.
\newblock \showarticletitle{A systematic literature review of servant
  leadership theory in organizational contexts}.
\newblock \bibinfo{journal}{{\em Journal of Business Ethics\/}}
  \bibinfo{volume}{113}, \bibinfo{number}{3} (\bibinfo{year}{2013}),
  \bibinfo{pages}{377--393}.
\newblock


\bibitem[\protect\citeauthoryear{Petersen}{Petersen}{2011}]%
        {isleanagile}
\bibfield{author}{\bibinfo{person}{K. Petersen}.}
  \bibinfo{year}{2011}\natexlab{}.
\newblock \showarticletitle{Is lean agile and agile lean? {A} comparison
  between two software development paradigms}.
\newblock In \bibinfo{booktitle}{{\em Modern software engineering concepts and
  practices: {A}dvanced approaches}},
  \bibfield{editor}{\bibinfo{person}{A.~Dogru} {and}
  \bibinfo{person}{V.~Bicer}} (Eds.). \bibinfo{publisher}{IGI Global},
  \bibinfo{pages}{19--46}.
\newblock


\bibitem[\protect\citeauthoryear{Von~Cranach}{Von~Cranach}{1986}]%
        {von1986leadership}
\bibfield{author}{\bibinfo{person}{M. Von~Cranach}.}
  \bibinfo{year}{1986}\natexlab{}.
\newblock \showarticletitle{Leadership as a function of group action}.
\newblock In \bibinfo{booktitle}{{\em Changing conceptions of leadership}},
  \bibfield{editor}{\bibinfo{person}{C.~Graumann} {and}
  \bibinfo{person}{S.~Moscovici}} (Eds.). \bibinfo{publisher}{Springer},
  \bibinfo{address}{New York}, \bibinfo{pages}{115--134}.
\newblock


\bibitem[\protect\citeauthoryear{Wheelan}{Wheelan}{2005}]%
        {wheelandev}
\bibfield{author}{\bibinfo{person}{S Wheelan}.}
  \bibinfo{year}{2005}\natexlab{}.
\newblock \bibinfo{booktitle}{{\em Group processes: {A} developmental
  perspective\/} (\bibinfo{edition}{2} ed.)}.
\newblock \bibinfo{publisher}{Allyn and Bacon}, \bibinfo{address}{Boston}.
\newblock
\showISBNx{0-205-41201-7}


\bibitem[\protect\citeauthoryear{Wheelan and Hochberger}{Wheelan and
  Hochberger}{1996}]%
        {wheelan}
\bibfield{author}{\bibinfo{person}{S. Wheelan} {and} \bibinfo{person}{J.
  Hochberger}.} \bibinfo{year}{1996}\natexlab{}.
\newblock \showarticletitle{Validation studies of the group development
  questionnaire}.
\newblock \bibinfo{journal}{{\em Small Group Research\/}} \bibinfo{volume}{27},
  \bibinfo{number}{1} (\bibinfo{year}{1996}), \bibinfo{pages}{143--170}.
\newblock


\bibitem[\protect\citeauthoryear{Williams}{Williams}{2012}]%
        {williams}
\bibfield{author}{\bibinfo{person}{L. Williams}.}
  \bibinfo{year}{2012}\natexlab{}.
\newblock \showarticletitle{What agile teams think of agile principles}.
\newblock \bibinfo{journal}{{\em CACM\/}} \bibinfo{volume}{55},
  \bibinfo{number}{4} (\bibinfo{year}{2012}), \bibinfo{pages}{71--76}.
\newblock


\end{thebibliography}

\end{document}